\documentstyle[epsf,12pt]{article}

\begin{document}
\epsfysize 3cm
\epsfbox{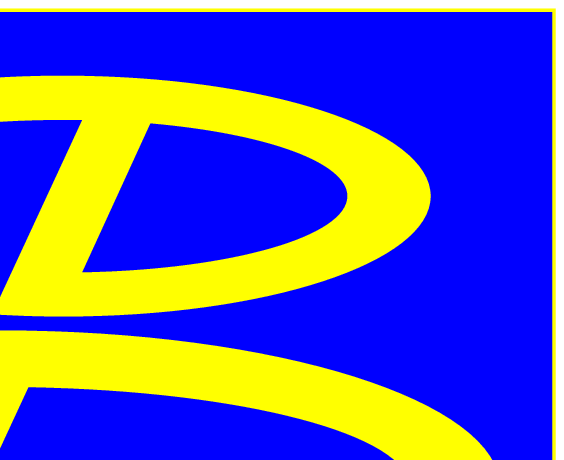}
\vskip -3cm
\noindent
\hspace*{4.5in}BELLE Preprint 97-5 \\
\hspace*{4.5in}KEK Preprint 97-102 \\
\begin{center}
\vskip 2cm
{\Large \bf
Monte-Carlo Simulation for an Aerogel \v Cerenkov Counter\footnote{
Submitted to Nucl. Instrum. Meth. {\bf A}.}
}\\
\vskip 1cm
R. Suda$^a$,
M. Watanabe$^b$,
R. Enomoto$^c$\footnote{Electric Mail Address: enomoto@bsunsrv1.kek.jp.},
T. Iijima$^c$,
I. Adachi$^c$,
H. Hattori$^d$,
T. Kuniya$^e$,
T. Ooba$^b$,
T. Sumiyoshi$^c$,
and Y. Yoshida$^f$\\
\vskip 1cm
{\it
$^a$Tokyo Metropolitan University, Tokyo 192-03, Japan\\
$^b$Chuo University, Tokyo 112, Japan \\
$^c$High Energy Accelerator Research Organization, KEK, Ibaraki 305, Japan\\
$^d$Chiba University, Chiba 260, Japan \\
$^e$Saga University, Saga 840, Japan\\
$^f$Toho University, Chiba 274, Japan\\
}
\end{center}
\newpage
\begin{abstract}
We have developed a Monte-Carlo simulation code for an aerogel \v Cerenkov
Counter which is operated
 under a strong magnetic field such as 1.5T.
This code consists of two parts: photon transportation inside aerogel 
tiles, and one-dimensional amplification in a fine-mesh 
photomultiplier tube.
It simulates the output photoelectron
yields as accurately as 5\% with  only a single free parameter.
This code is applied to simulations for a
B-Factory particle-identification system.
\\


keywords : aerogel, Fine-Mesh photo multiplier tube, Monte-Carlo
\end{abstract}
\newpage
\section{Introduction}

Particle identification (PID), particularly the identification of charged
pions and kaons, plays an important role in CP-violation
studies in B-Factory experiments\cite{ref:rossia}.
In the BELLE experiment at the KEK B-factory(KEKB), 
a threshold silica aerogel
\v Cerenkov counter (ACC) will be used to extend the
momentum region of the PID capability
beyond the reach of $dE/dx$ and time-of-flight
(TOF) measurements\cite{ref:tdr}.




Silica aerogels (aerogels) are a colloidal form of glass, in which
globules of silica are connected in three-dimensional networks
with siloxan bonds\cite{ref:adachi,ref:sahu}.
Many high-energy and nuclear-physics experiments have used aerogels
instead of pressurized gas for \v Cerenkov counters.

In the BELLE experiment, fine-mesh photomultiplier tubes
(FM-PMTs) will be used for the readout of \v Cerenkov photons emitted from
aerogel\cite{ref:fractal,ref:beaune}.
This was only choice at the design period
for this kind of counter operated under a strong
magnetic field, such as 1.5T.


We have developed a Monte-Carlo simulation for the aerogel counter,
which includes photon transportation in the aerogel tiles and response 
of a phototube.
One of the key issue in this simulation is the
parameterization of absorption length, which is obtained by a careful
comparison of beam test data and the prediction from the simulation.
We describe a Monte-Carlo simulation of this detector system 
in this paper.
In the next section we introduce our experimental setup of  
the threshold \v Cerenkov counter. 
The purpose of this Monte-Carlo project is also described.
The framework of this Monte-Carlo code
is described in the third section.
The simulation of fine-mesh phototubes is described in section 4,
and photon transportation inside aerogel tiles is shown in section 5.
In section 6, a simple model for aerogel absorption is introduced.
The results on this model is described in section 7.
In section 8, we discuss on the wavelength dependence of the
light absorption.
The conclusion is given in section 9.

\section{Threshold \v Cerenkov Counter with a Fine-mesh 
Photomultiplier Tube Readout}

We will use threshold \v Cerenkov counters to identify
charged kaons in the momentum region 1-4 GeV.
The configuration of the aerogel counter system of the BELLE detector
is shown in Figure \ref{fig:side_view}.
It should be noted that
the KEK B-factory is an asymmetric $e^+e^-$ collider
with  energies of 3.5 $\times$ 8 GeV and the produced particles
from the B meson decay
are Lorentz boosted to the $e^-$-beam direction(z). 
Since the momentum range of the produced pions and kaons
changes as the emitted polar angle changes in a laboratory system,
we employ the different refractive indices (n) of aerogel tiles
corresponding to the different polar angle.
In the endcap region, however, $n$=1.03 was selected for
$B$-flavor tagging purpose\cite{ref:aleksan}.

The typical arrangement of the counter is 
shown in Figure \ref{fig:single_module}.
The size of an aerogel tile is approximately
12 $\times$ 12 $\times$ 2.4 cm$^3$. The single counter
contains 4, 5, or 6 layers of them with a total thickness 
of 10-14.5 cm.
The \v Cerenkov photons are readout by one or two fine-mesh 
photomultiplier tubes
(FM-PMT). 
In order to reduce effects of a magnetic field of 1.5T in the $z$ direction,
we selected the FM-PMT.
We will use three types of FM-PMTs of 2, 2.5, and 3
inches in diameters. The 
effective diameters of these FM-PMTs are 39, 51, and 64 mm.
The inner side of the counter is covered with a white reflector 
(Goretex\cite{ref:goretex}).

The dominant scattering mechanism of the \v Cerenkov photon inside 
the aerogel tile is considered to be Rayleigh scattering. 
There is almost no understanding about the absorption mechanism of the 
light inside aerogel. 
The main purpose of developing this Monte-Carlo code is to understand this
mechanism, or in other words, to construct a reliable model.


The amplification mechanism of the FM-PMT in a strong magnetic field is also
poorly understood. 
Since the B-factory is a high-statistics experiment,
an accurate Monte-Carlo simulation is necessary for these PMT's responses.

In summary, we would like to develop a Monte-Carlo 
simulation with an accuracy of
a few percent level. 
It should simulate such effects as position dependence and
edge effect.

\section{Framework}

The typical scattering length of aerogel is a few cm at $\lambda=$400nm
\cite{ref:sahu}.
Therefore, \v Cerenkov photons undergo many scatterings 
(including absorptions and
reflections) before reaching photocathodes.
It is very difficult to understand this situation 
analytically. The Monte-Carlo simulation
may be the only calculation choice.

In coding the Monte-Carlo simulation, we used C++ language.
A part of GEANT3.21\cite{ref:geant} was used for the geometry and 
display parts.

The flow chart of this Monte-Carlo simulation
(hereafter, referred as ``ACC++")
is shown in Figure \ref{fig:flow_chart}.
Each box corresponds to 
a major process in the program, which is written as
``class" in C++.
The vertical arrows indicate the direction of the data flow.
The display and geometry parts are common utilities.
At first the hit points by the charged particles inside
the aerogels are recorded. 
\v Cerenkov photons are emitted at the above-mentioned 
hit points
according to the velocity of the charged particles
and the refractive index of medium.
Then, the photon transport part takes care of photon transportation
to the phototube window surface. Finally, the FM-PMT part amplifies
the photoelectrons and the ADC counts are obtained.
For data management, we used the CLHEP library\cite{ref:clhep}.

\section{Fine-mesh Photomultiplier-Tube Simulation}

\subsection{Structure of a Fine-mesh PMT} 

A sectional view of the FM-PMT is shown 
in Figure \ref{fig:fine_mesh}.
Each PMT has a borosilicate glass window, a bialkari photocathode, 
19 fine-mesh dynodes and an anode. 
The diameters of the effective area($\phi$), equal to the diameters of 
the dynodes, are 39, 51, and 64mm for 2'', 2.5'', 
and 3'' PMTs, respectively.
The cathode-to-anode distances(L) are 20, 20, and 23mm 
for 2'', 2.5'', and 3'' PMTs, respectively.
The FM-PMT used for the BELLE ACC has the following 
improved performance.
The average quantum efficiency(QE) of the photocathode is 25\% at 400nm
wavelength, improved for recent products \cite{ref:hpk}.
A finer mesh than that of 
conventional products is used to improve the gains in
the magnetic fields \cite{ref:beaune}.
The optical opening of the mesh is about 50\%\cite{ref:hakamata}.
\subsection{Simple Model of Amplification}

The first step of the simulation is conversion of incident photons to 
photoelectrons(p.e.).
This conversion is simulated while taking into account the 
wavelength dependence of the quantum efficiency($QE$).
Numerical data of the wavelength dependence, obtained from 
the manufacturer, are used.
The absolute value of $QE$ for each PMT is determined by the cathode blue 
sensitivity, denoted as $Skb$\cite{ref:hpk}.
The absolute $QE$ at 400nm wavelength($QE$(400nm)) can be calculated as
\begin{equation}
QE(400{\rm nm})=0.026 \times Skb 
\end{equation}
using the data obtained from the manufacturer.
The $QE$ does not depend on the diameter of the PMT. 
The average $Skb$ value is 9.7.

The second step is to simulate the 
amplification in fine-mesh dynodes. A simple model is
schematically shown in Figure \ref{fig:amp_model}.
For each dynode stage, some incident electrons pass through holes of 
a fine-mesh dynode, and others hit the dynode wires, 
followed by the emission of 
secondary electrons.
The probability of passing through holes is proportional to the
optical opening(the ratio of the hole area to the total area) of the
mesh, $f(\sim 0.5)$.
Some of the secondary electrons pass through the holes, 
and others are re-absorbed by the same wire.
The re-absorption probability is dependent on the hit position,
and it is higher when electrons hit the mesh at the top of wires,
compared to the case when they hit it at the edge.
That probability increases at higher magnetic fields.
This dependence is considered in the simulation by introducing a blind 
region, where all of the secondary electrons are assumed to be
re-absorbed.
The ratio of the blind region area to the total dynode area is
represented by a parameter $b$ in the simulation.
The effective gain per one stage of a dynode(${\delta}^{eff}$) is then
parameterized as
\begin{equation}
{\delta}^{eff} = f + {\delta}^{hit} (1-f-b) ,
\label{eq:gain}
\end{equation}
where ${\delta}^{hit}$ is the number of secondary electrons passing
through holes, whose distribution is assumed to be a Poisson in the
simulation.
The quantity ${\delta}^{hit}$ is determined so that the total gain for 
the 19 stages($G$) is reproduced.
Since the gains of the first three and the last two dynodes 
of our FM-PMTs are
improved by a factor of $k$\cite{ref:hakamata}, we obtain
\begin{equation}
k^5\times ({\delta}^{eff})^{19} = G .
\label{eq:totalgain}
\end{equation}
Here, $\delta^{eff}$ is typically around 2.5\cite{ref:hakamata}.

In a simulation of the pulse height for each incident photoelectron,
the above process is repeated until the sixth dynode.
It is shown later that a calculation up to the sixth dynode is 
sufficient to
reproduce the pulse-height spectrum.

\subsection{Single-photoelectron Spectrum}

Using the method described in section 4.2, a pulse-height spectrum for
single-photoelectron events is simulated, as shown in Figure
\ref{fig:single_pe}-(a). 
That is compared to the experimental data shown in Figure
\ref{fig:single_pe}-(b).
The experimental data were taken by using a pulsed laser\cite{ref:PLP}.
The number of events in the pedestal region is more than 90\% of
the total events.
Therefore, most of the signal events are single-photoelectron events.

The simulation reproduces the experimental data well.
The single-photoelectron spectrum of FM-PMTs has no 
characteristic peak, 
in contrast with that of line-focus PMTs.
Only a fractal structure is observed in the spectrum,
the interpretation of which is described in reference \cite{ref:fractal}.

\subsection{Multi-photoelectron spectrum}

As noted in section 4.3, a FM-PMT has no peak corresponding to 
a single photoelectron
in the spectrum.
This deteriorates the pulse-height resolution of a
multi-photoelectron spectrum, compared  to the resolution determined
by the original Poisson distribution of incident photoelectrons.

Figure \ref{fig:deterio_mc} shows a comparison of the incident
photoelectron distribution(a) and the simulated spectrum at the anode
for corresponding events(b).
To quantitate the deterioration, we define the excess noise
factor($ENF$) as
\begin{equation}
ENF= \frac{\mu_{input}}{\mu_{eff}} ,
\end{equation}
where $\mu_{input}$ is the average number of input photoelectrons and
$\mu_{eff}$ is the effective number of output photoelectrons.
In the present work, the output spectrum is fit to an asymmetric
Gaussian with two width parameters ($\sigma_1$ and $\sigma_2$),
each representing the width of the 
lower and higher sides, respectively, as
shown in Figure \ref{fig:deterio_mc}-(b).
Then $\mu_{eff}$ is calculated as
\begin{equation}
\mu_{eff} = \left( \frac{\overline{n}}{\sigma_1} \right)^2 \,,
\end{equation}
where $\overline{n}$ is the mean of the spectrum (mean ADC count).
In the example shown in Figure \ref{fig:deterio_mc}, $\mu_{input}$ is
20.0 and $\mu_{eff}$ is 9.64. Therefore, $ENF$ is 2.07.

Figure \ref{fig:deterio_exp} shows the measured $ENF$ for 2'',
2.5'', and 3'' PMTs.
The tested PMTs were illuminated by the same
412nm pulsed laser light.
The number of incident photons was calibrated using a reference PMT,
that is a 2'' line-focus type PMT (Hamamatsu R329-05S), which was well
calibrated by the manufacturer.
The average number of photoelectrons($\mu_{input}^{data}$) was deduced from
the average ADC channel($\overline{q}$) and gain.
The gain was obtained from the mean ADC channel for a
single-photoelectron event($q_0$).
Taking into account of the difference in the quantum efficiency
between the reference PMT($QE$(ref)) and the tested FM-PMT($QE$(FM)),
the number $\mu_{input}^{data}$ was obtained as
\begin{equation}
\mu_{input}^{data} = \frac{QE({\rm FM})}{QE({\rm ref})} 
\frac{\overline{q}}{q_0}.
\end{equation}
The quantum efficiencies ($QE$(FM) and $QE$(ref)), measured by the
manufacturer have about a 10\% ambiguity, giving about a 14\% error in
the measured $ENF$\cite{ref:hakamata}.
The measured $ENF$ is 1.99 on the average.
The agreement between the simulation and the data is satisfactory.

\subsection{Effect of a Magnetic Field on the Resolution}

The electron trajectories are influenced by magnetic fields.
In the simulation, the effects of magnetic fields are treated as follows.

On the microscopic scale, electrons spiral along the magnetic field.
At 1.5T, the Lamor radius of electrons for a typical
energy of 6eV is 5.5$\mu $m, which is comparable to the mesh width.
Therefore, the 
probability for the re-absorption of secondary electrons by mesh
wires increases.
This effect can be regarded as an increase in the blind region($b$) on
the mesh, which is shown in Figure \ref{fig:amp_model}, not only as a
decrease in ${\delta}^{hit}$. 
The increase in $b$ affects the pulse-height resolution, while a
decrease in ${\delta}^{hit}$ affects the total gain.

On the macroscopic scale, electrons move parallel to the direction of the
magnetic field.
Therefore, some electrons can not reach the anode when the PMT axis is
inclined with respect to the magnetic field.
The fraction of such electrons is determined by the diameter of
the photocathode($\phi$) and the cathode-to-anode distance(L).
In the simulation, the 
position of secondary electrons on the anode plane
is calculated from information about the positions of electrons at the
cathode and the field direction.
Then, in case that an electron is in the effective area of the anode,
the amplification process is simulated.
 
Table \ref{tab:MCvsDATA} shows the ratio of the effective number of
photoelectrons with magnetic fields of 1.5Tesla ($\mu_{eff}(1.5T)$) to
that without magnetic fields ($\mu_{eff}(0T)$), obtained from our
simulation and a measurement for each PMT size.
The results are shown for $\theta =0^{\circ}$ and $35^{\circ}$, where
$\theta$ is the angle between the PMT axis and the fields. 
The parameter $b$ is adjusted to reproduce the experimental data for
each angle.
In the case of $\theta =0^{\circ}$, the experimental data can be
reproduced well with $b = 0.15$.
In the case of $\theta =35^{\circ}$, the results of the simulation are 
shown for two cases, $b = 0$ and $b = 0.15$.
A better agreement between the data and the simulation is obtained
when only the macroscopic effect is considered ($b = 0$).
Therefore, it is indicated that the macroscopic effect dominates for
$\theta = 35^{\circ}$ case.
The dependence of the parameter $b$ on the angle $\theta$ can be
explained by a change of the mesh width projected along the magnetic
field direction.

\section{Photon Transport inside Aerogel Tiles}
\subsection{Algorithm}

A charged particle is transported inside aerogels
by a 1-mm step.
\v Cerenkov photons are radiated along the 
charged-particle's trajectory.
The number of photons ($N_{p}$) is given by the Frank-Tamm
equation \cite{ref:frank},
 \begin{equation}
  \frac{dN_{p}}{dE} = (\frac{\alpha}{\hbar c})L z^2 \sin^2 \theta_c,
 \end{equation}
where $\alpha$ is the fine-structure constant, 
$L$ the thickness of the radiator, 
$z$ the charge of an incident particle,
$\theta_c$ the \v Cerenkov angle, 
and $E$ the energy of a radiated photon.
The \v Cerenkov angle ($\theta_c$) for a particle with velocity
$\beta c$ in a medium with index of refraction ($n$) is
 \begin{equation}
  \theta_c = \cos^{-1} (\frac{1}{\beta n}).
 \end{equation} 

A radiated photon is transported according to the algorithm
shown in 
Figure \ref{fig:algorithm}.
In the photon-transport part,
we first obtain information about photon, i.e., the wavelength,
radiated position, and radiated direction
from the \v Cerenkov-radiation part.

The \v Cerenkov photon is transported to the next position in the
following way.
The mean transport length ($\Lambda$) is defined as
   \begin{equation}
     \frac{1}{\Lambda(\lambda)} = \frac{1}{\Lambda_{scat}(\lambda)} + 
                                      \frac{1}{\Lambda_{abs}(\lambda)} ,
   \end{equation}
where $\Lambda_{scat}$ and $\Lambda_{abs}$ are the scattering length,
and absorption length respectively, in the aerogel.
They are calculated based on the spectrophotometer
measurement, which is described in section 5.2.1.
The transport length is calculated using
the exponential probability function (Poisson distribution).
It is then judged that the position after transport
is (i) outside the counter box or (ii) inside with the information 
from the Geometry part,
in which we used GEANT 3.21 (only geometry routines)\cite{ref:geant}.
(i) If the position is out of the box, 
the photon is traced to the inner wall of
the box.
We then judge whether the position is on the FM-PMT surface or
diffuse reflector surface (Goretex \cite{ref:goretex}).
If the position is on the FM-PMT surface, the photon information
(the wavelength and the position of hit) is given to FM-PMT simulation
 part and the photon transport is finished.
If the photon hits a reflector surface, we judge whether the photon is
absorbed or reflected by Goretex based on a spectrophotometer
measurement of the Goretex reflectivity, which is
described in section 5.2.2.
If the photon is absorbed by Goretex, the photon transport is finished.
If the photon is reflected by Goretex (diffuse reflection),
 the photon is transported again.
(ii) If the position after transportation is in the counter box, 
it is judged that the photon is absorbed or scattered by aerogel,
according to the following ratios:
 \begin{eqnarray}
  P_{abs} = \frac{\Lambda_{scat}}{\Lambda_{scat} + \Lambda_{abs}} , \\
  P_{scat} = \frac{\Lambda_{abs}}{\Lambda_{scat} + \Lambda_{abs}} ,
 \end{eqnarray}
where the $P_{abs}$ and $P_{scat}$ are the probability of absorption and the
scattering of the aerogel, respectively.
If the photon is absorbed by aerogel, the photon transport is finished.
If the photon is scattered by aerogel,
we randomize the direction according to the 
Rayleigh-scattering 
formula and repeat the above-mentioned transportation again.
\subsection{Input Parameters}
\subsubsection{Spectrophotometer Measurement}

We use the spectrophotometer\cite{ref:monochrometer} in order to
measure the transmittance of aerogel and reflectivity of Goretex as a
function of the wavelength(250 $\sim$ 800 nm).
The transmittance ($T$) is defined as $I/I_0$, where $I$ and $I_0$
are the photon fluxes at the photodetector 
with and without aerogels, respectively.
The thickness of a typical aerogel tile is 24mm.
The spectrophotometer has a finite solid angle at
the entrance of the photo detector. 
A part of scattered photons is detected by the photodetector.
Therefore, the transmittance 
($T$) is
 \begin{eqnarray}
  T \ne 1 - A, \\
  T \ne 1 - A - S,
 \end{eqnarray}
where $A$ is the absorption and $S$ is the scattering.
In order to estimate the contribution
from $S$ and $A$ to $T$, we took the geometry
of the photospectrometer and other necessary information
into account for our Monte-Carlo (ACC++) and carried out
a simulation.
The details are discussed in section 5.2.3.
A simulation display of the photospectrometer is shown in
Figure \ref{fig:spectro}. 
The lines are the photon trajectories. 
The circle is the aperture of the
photodetector (8mm$\phi$). 
\subsubsection{Goretex Reflectivity}

It is not possible to measure the absolute reflectivity of a material
directly with the spectrophotometer.
We therefore measure the relative reflectivities of samples to a
reference sample whose absolute reflectivity is known.
We used an NIST traceable standard reflector (Spectralon) which was
well calibrated by a company\cite{ref:spectron}.
Then, the absolute reflectivity of Goretex was calculated.
Figure \ref{fig:goretex_ref} shows the measured absolute reflectivity
of the Goretex as a function of the wavelength.
The reflectivity of Goretex is better than 93\% over the range of
the measurement.
It shows a better reflectivity in the short-wavelength region,
i.e., a better acceptance for \v Cerenkov light (UV light).
\subsubsection{Rayleigh Scattering}
We input two parameters ($\Lambda_{abs}$ and $\Lambda_{scat}$)
in the simulator.
Initially, we suppose that the transmittance is only due to
the Rayleigh-scattering effect.
We consider the absorption and scattering parameter, 
as expressed below
by Equations \ref{eq:abs_inf} and \ref{eq:scat_tr}, 
and input them to
the simulator:
 \begin{eqnarray}
  \Lambda_{abs} = \infty,\label{eq:abs_inf}\\
  \Lambda_{scat} = \frac{-d}{\ln T},\label{eq:scat_tr}
 \end{eqnarray}
where $d$ is the thickness of the samples.
Although the simulated transmittance shows almost the 
same shape as the experimental
transmittance, the absolute value of the simulated
data points are higher than the experimental data (i.e., more transparent).
 This indicates that
the scattered events within the finite solid angle of the photodetector
and
the absorption events are taken into account for the transmittance.
We must take these factors into account.
Then, the aperture factor ($f_a$) is 
introduced in Equation \ref{eq:aperture}
and the definition of $\Lambda_{scat}$ is modified to
 \begin{equation}
  \Lambda_{scat} = f_a \times \frac{-d}{\ln T}.
  \label{eq:aperture}
 \end{equation}
To evaluate the aperture factor ($f_a$), a simulation was carried out.
The obtained value for $f_a$ is 0.55.
Figure \ref{fig:aperture} shows the 
transmittance data and simulation result,
which agree with each other.
In other words, the photospectrometer measurement of our
aerogels suggests that there is no, or very small, absorption.

\section{Model for Absorption}
Empirically, it was known that the transmittance obtained
by the spectrophotometer was well fitted with the
two terms:
 \begin{eqnarray}
    \Lambda_{scat} = a\lambda^4,~{\rm and }~ \Lambda_{abs} = b\lambda^2,
    \label{eq:slac}
 \end{eqnarray}
where $\lambda$ is the wavelength and $a$ and $b$ are free parameters.
The $\lambda^2$ term was empirically introduced in order to improve
the fitting of the transmittance spectra.
In the reference\cite{ref:slac}, they assumed that 
the $\lambda^4$ term is the
Rayleigh scattering, and that the $\lambda^2$ term is the absorption.
The results ($n$=1.010) are shown in Figure \ref{fig:trans_fit}
for our aerogel.

The typical scattering length was calculated to be
$\sim 2.5$ cm at 400 nm wavelength, and the absorption
length was fitted to be 10-times the scattering length.
When we input parameters $\Lambda_{scat}$ and $\Lambda_{abs}$ 
into our Monte-Carlo simulation (ACC++),
we obtained only $\sim$ 2 photo-electrons (p.e).
On the other hand, an experiment using 3.5 GeV $\pi^-$ beam
was carried out for our counter box (Figure \ref{fig:single_module});
we obtained $\sim$20 p.e.,
which is inconsistent with the above assumption.
Therefore, we conclude
that $b\lambda^2$ is not due to absorption,
 and that the absorption length
is greater than 100-times the scattering length.

In order to estimate the magnitude of the absorption length,
we make a simple model that the absorption length is proportional
to the scattering length, 
i.e., we assume the same wavelength dependence
as that of the Rayleigh scattering,
 \begin{equation}
  \Lambda_{abs} = \alpha \times \Lambda_{scat},
 \end{equation}
where $\alpha$ is a free parameter.
The $\alpha$ value was 
determined by comparing the experimental data and
simulation results as described in the next section.
We then input these two parameters 
($\Lambda_{abs}$ and $\Lambda_{scat}$)
into ACC++.
Figure \ref{fig:test} shows a simulation display of ACC++
for the photon transportation inside the aerogel tile with the
assumption $\alpha$=100.
The other type of absorption model can be found, for example, in 
reference\cite{ref:arisaka}.
\section{Comparison between the Beam-Test Results and Simulation}

\subsection{Determination of $\alpha$ by $\chi ^2$ Minimization}

We carried out a $\chi^2$ minimization 
by comparing the $\mu_{eff}$ value
of the experimental data and that of the Monte-Carlo data
while changing the free parameter ($\alpha$).
The experimental data were obtained in the beam test with 
the 3.5-GeV $\pi^-$ at the $\pi2$ beam line of
the KEK PS.
The incident position of the $\pi^-$'s is the center of the box.
Figure \ref{fig:chisq} shows this result.
From the minimum of this curve we 
obtained $\alpha = 408.1 \pm ^{68.3}
_{58.5}$ which corresponds to an absorption length ($\Lambda_{abs}$) 
of about 5.4 m at 400 nm.
The $\alpha$ values differ (by a factor of $\sim$2)
for different samples of aerogels having the same refractive
index and a different refractive index.
There are no clear dependence on transmittance nor refractive indices.
We, therefore, must obtain this value for each counter box, for example
using $e^+e^-\rightarrow \mu^+\mu^-$ events
in the real B-factory experiment.
\subsection{Position Dependence and Accuracy}

With the best-fitted $\alpha$ value we 
compared the beam data and the Monte-Carlo prediction for the
 position dependence
of light yields.
The experimental data of $\mu_{eff}$ were obtained for 
the 3.5-GeV $\pi^-$.
The data were taken at various positions
of the incident beam. We selected 23 incident positions.
The coordinate system of the beam incident position 
and direction are defined in Figure
\ref{fig:pos_inc}.
The beam direction is (0, 0, 1) for all data.
A comparison with the Monte-Carlo data are shown in Figures
\ref{fig:pos_dep}:
(a) at y=0 plane, (b) y=2cm, (c) y=3.5cm, and (d) y=4cm.
The hatched areas are the Monte-Carlo predictions.
These errors are statistical.
The disagreements between the experiment and Monte-Carlo
values are around
the 5\% level. 
We therefore consider the accuracy of our Monte-carlo 
simulation to be 5$\%$.

\subsection{Miscellaneous Distributions}

With the best-fitted value of the parameter $\alpha$, 
we derived the distributions of the number of
scatterings inside the aerogel which our
Monte-Carlo simulation predicts.
They are shown in 
Figure \ref{fig:trans_tot}.
The mean number of reflections on the Goretex is $\sim$ 11-times,
that of the scattering in aerogel is $\sim$ 29-times,
and mean total transport length is $\sim$ 83 cm.
43$\%$ of initial radiated photons in aerogel reach the PMT surface,
32$\%$ are absorbed by Goretex,
and 25$\%$ are absorbed by aerogel.
This Monte-Carlo code is very useful 
in designing the shape of the counter box
for future experiments.

\section{Wavelength Dependence of Absorption}
As noted in section 7.2, our Monte-Carlo simulator is able to reconstruct
beam data with 5$\%$ accuracy after fitting the $\alpha$ value.
For a further study to evaluate the
wavelength dependence of the absorption
length($\Lambda_{abs}$), we carried out experiments 
with several types of optical filters (sharp-cut filters
\cite{ref:filter}).
These filters cut the light with a 
shorter wavelength than threshold values.
Here, we define the filter's name, whose threshold wavelength is 300nm,
300nm-filter.
We used eight types of filters,
whose threshold wavelengths are 300nm, 360nm, 400nm, 460nm, 500nm, 560nm,
600nm, and 700nm. The transmittance of each filter was measured with
spectrophotometer as a function of the wavelength and input to ACC++.

The beam-test setup and evaluation procedure of the $\alpha$ value were
the same as those 
mentioned in section 7.1. The refractive index of aerogel
was $n = 1.013$. We used a 3" FM-PMT with a 2" mask, 
because the diameter of
the filters were 2".
The filter was set between the PMT surface and the aerogel tiles. 
After fitting beam data with the simulation results,
we obtained the best $\alpha$ value for each
filter setup, $\alpha = 302.7 \pm^{7.0}_{6.8}$ for the 300nm filter, 
$\alpha = 596.8 \pm^{32.5}_{30.8}$ for the 
360nm filter, $\alpha = 3104.5
\pm^{761.4}_{611.5}$ for the 400nm filter and $\alpha$ diverges for
the longer-wavelength filters. 
These are plotted in Figure \ref{fig:filter} as
a function of the threshold wavelength of the filter.
From these $\alpha$ values, we can conclude 
that $\Lambda_{abs}$ is not proportional
to $\lambda^4$. 
The absorption is significant
only in short-wavelength range (around 300nm), 
but not in the longer wavelength region ($>$400nm).

Another absorption model, which assumes 
$\Lambda_{abs} = \exp(\beta \lambda)$,
is found in the reference \cite{ref:arisaka}.
Here $\beta$ is a free parameter.
An attempt to evaluate $\beta$ was carried out using
the $\chi^2$ minimization with the center incidence data.
Then we looked at the wavelength dependence (experimental data
with the sharp-cut filters) of the $N_{pe}$.
In the result, we find that $\beta$ is still dependent on
the threshold wavelength of filter. The $\beta$ value increases at
longer wavelength.
This again suggests that the absorption only happens at short wavelength
such as 300nm. 

\section{Conclusion}

We have developed a Monte-Carlo Simulation for an aerogel \v Cerenkov counter
under a strong magnetic field.
This code consists of two parts: photoelectron amplification
in a fine-mesh phototube, and
photon transportation inside aerogel tiles.

In the former, we have developed a simple one-dimensional 
fractal-amplification model. 
It successfully reproduces the deterioration
of the resolution and the single-photoelectron response in a fine-mesh
phototube.

In the latter, we have proven that recently produced aerogels have a significantly
longer absorption length, such as longer than a few meters.
Here, a simple model agrees with the experimental
observations. This model only uses a single free parameter.

With this Monte-Carlo simulation code, we can simulate effective
photoelectron yields within 5\% ambiguities for various shape
of counters, whose geometry can be defined in the GEANT framework. 
This program will be used for a particle-identification system for
the KEK B-Factory, BELLE experiment.

\section*{Acknowledgment}

We would like to thank the BELLE collaboration of the KEK B-Factory
for its help in this project.
We would especially like to thank Prof. N. Katayama (KEK) for coding
this program in C++ language.
We also appreciate the authors of GEANT 3.21 and CLHEP.
We are grateful to Prof. T. Hirose (Tokyo Metropolitan Univ.),
Prof. S. Matsumoto (Chuo Univ.), and Prof. A. Murakami (Saga Univ.)
for valuable comments on this manuscript.

The aerogels were developed under collaborative research
between Matsushita Electric Works Ltd. and
KEK.
We would finally like to thank Hamamatsu Photonics K.K. for developing
the fine-mesh photomultiplier tubes.

\newpage\section*{Table \ref{tab:MCvsDATA}, R.Suda et al., NIM-A} 
\begin{table}[h]
\caption{
Ratio of the effective number of photoelectrons with magnetic fields of
1.5 Tesla ($\mu_{eff}(1.5T)$) to that without magnetic
fields ($\mu_{eff}(0T)$) for 2'', 2.5'', and 3'' PMTs
for our simulation(MC) and experimental data(Data).
The detail on $\mu_{eff}$ can be found in the text.
}
\label{tab:MCvsDATA}
\begin{center}
\begin{tabular}{cc|cc|cc|cc}
\hline
\hline
\multicolumn{2}{c|}{} & \multicolumn{2}{c|}{2''} 
& \multicolumn{2}{c|}{2.5''} & \multicolumn{2}{c}{3''}\\
\hline
$\theta =0^{\circ}$ & MC & \multicolumn{2}{c|}{0.72}
&\multicolumn{2}{c|}{0.70}&\multicolumn{2}{c}{0.70} \\
 & Data & \multicolumn{2}{c|}{0.76}
&\multicolumn{2}{c|}{0.75}&\multicolumn{2}{c}{0.74} \\
\hline
$\theta =35^{\circ}$ & MC & 0.62 & 0.47 & 0.65 & 0.39 & 0.64 & 0.34 \\
                     &    & (b=0) & (b=0.15) & (b=0) & (b=0.15) & (b=0)
& (b=0.15) \\
  &  Data & \multicolumn{2}{c|}{0.59}& \multicolumn{2}{c|}{0.69}
&\multicolumn{2}{c}{0.67} \\
\hline
\hline
\end{tabular}
\end{center}
\end{table}

\begin{figure}[h]
\epsfbox{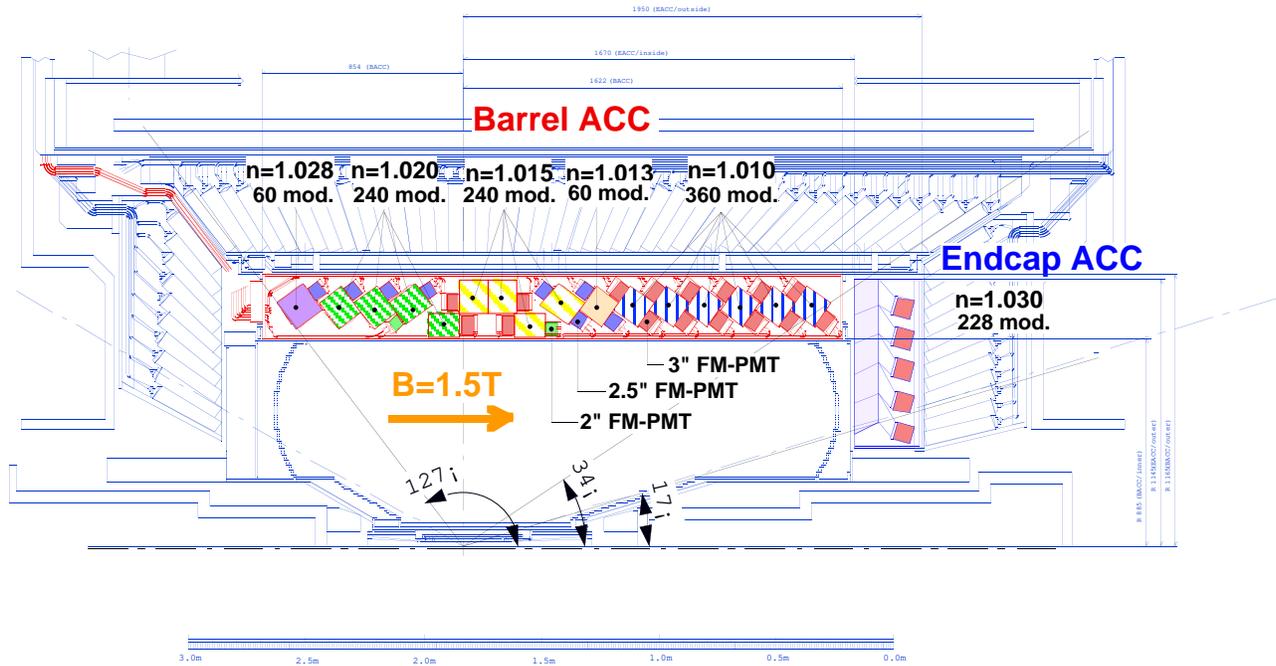}
\caption{
BELLE aerogel \v Cerenkov-counter system (ACC) and BELLE detector.
}
\label{fig:side_view}
\end{figure}

\newpage\section*{Figure \ref{fig:side_view}, R.Suda et al., NIM-A}

\begin{figure}[h]
\epsfbox{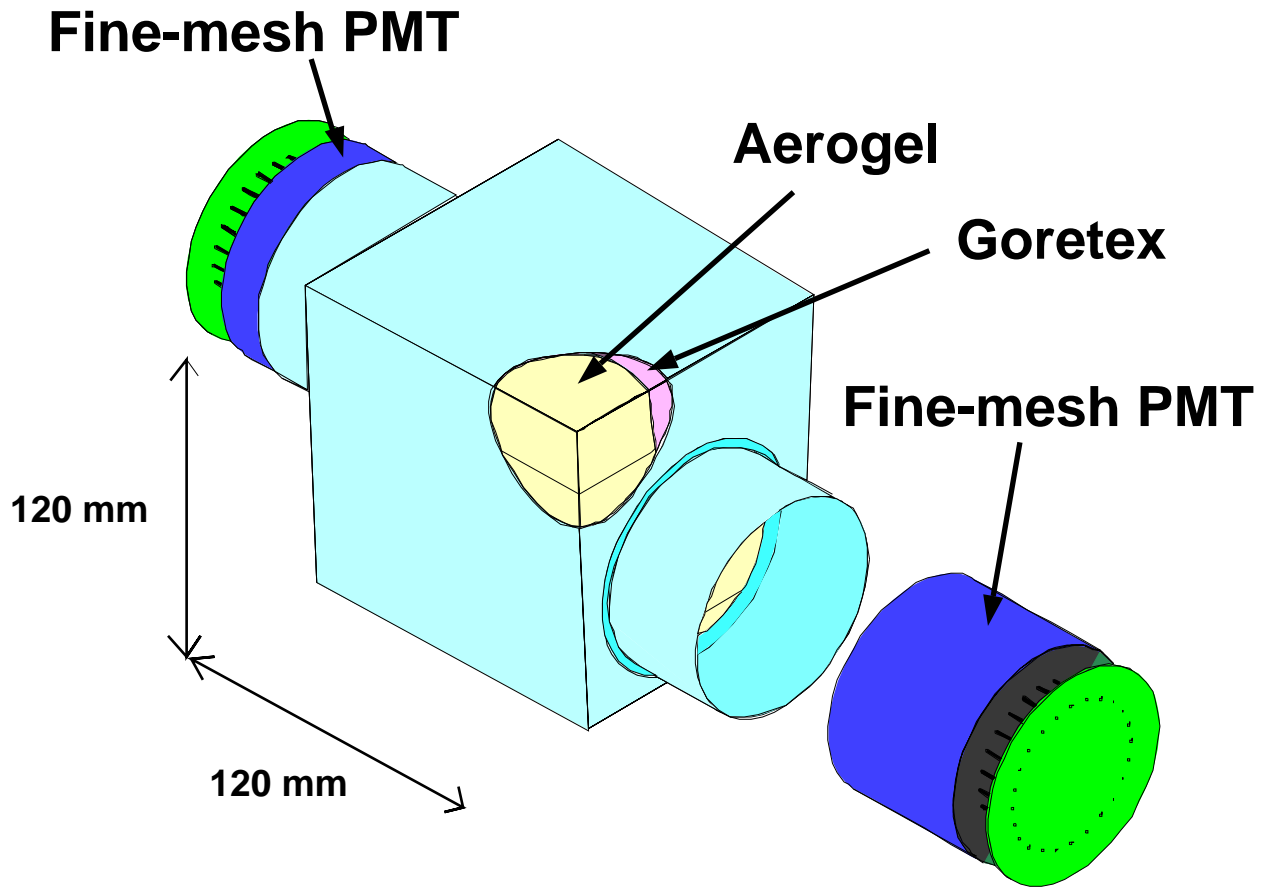}
\caption{
Typical form of the aerogel \v Cerenkov counter. 
}
\label{fig:single_module}
\end{figure}

\newpage\section*{Figure \ref{fig:single_module}, R.Suda et al., NIM-A}

\begin{figure}[h]
\epsfbox{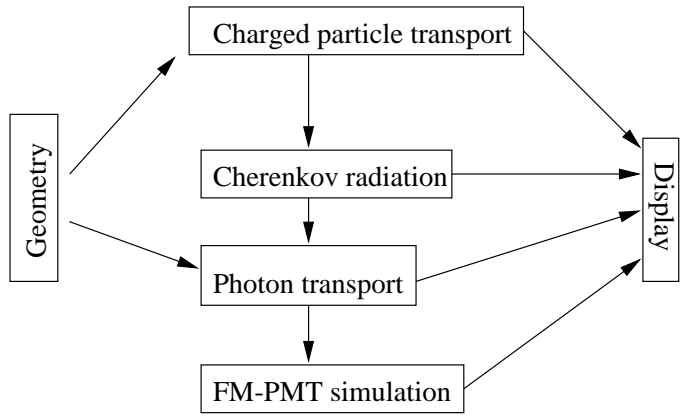}
\caption{Flow chart of 
the Monte-Carlo simulation code
(ACC++).}
\label{fig:flow_chart}
\end{figure}

\newpage\section*{Figure \ref{fig:flow_chart}, R.Suda et al., NIM-A}

\begin{figure}[h]
\epsfysize 10cm
\epsfbox{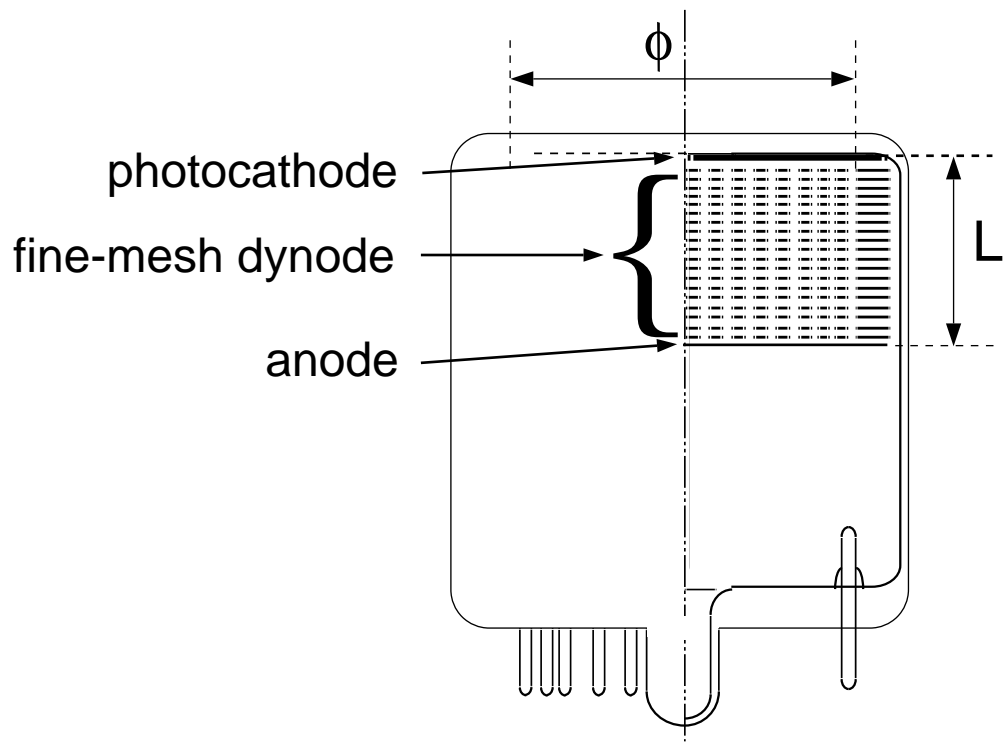}
\caption{Sectional view of FM-PMT.}
\label{fig:fine_mesh}
\end{figure}

\newpage\section*{Figure \ref{fig:fine_mesh}, R.Suda et al., NIM-A} 

\begin{figure}[h]
\epsfysize 10cm
\epsfbox{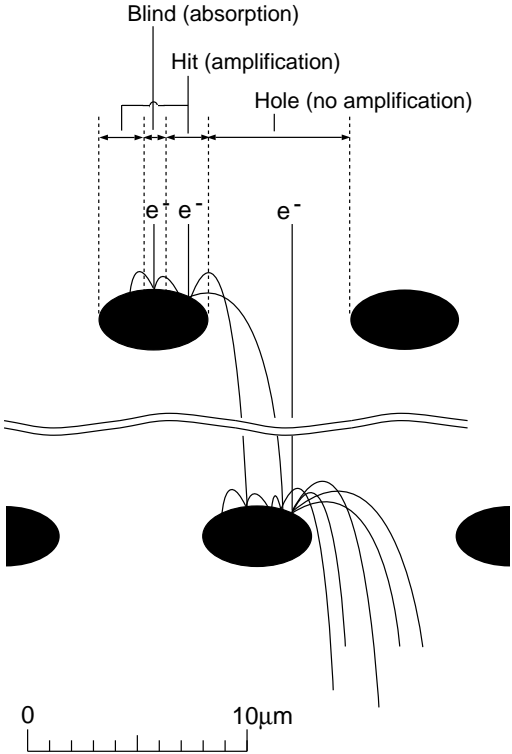}
\caption{Model of the amplification
by fine-mesh dynodes.}
\label{fig:amp_model}
\end{figure}

\newpage\section*{Figure \ref{fig:amp_model}, R.Suda et al., NIM-A} 

\begin{figure}[h]
\epsfysize 10cm
\epsfbox{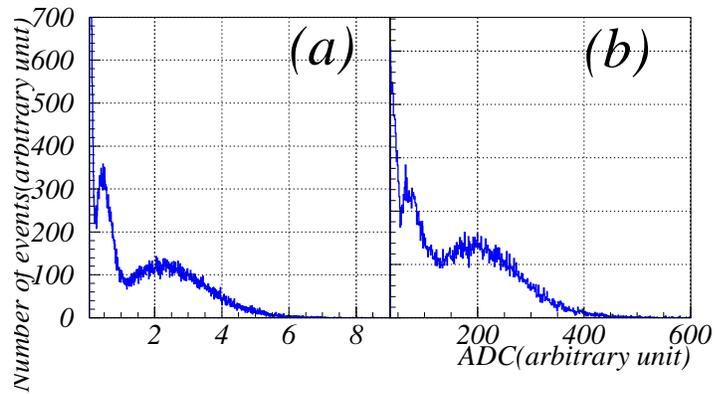}
\caption{Single photoelectron spectra: (a) simulated spectrum and
(b) experimental data.}
\label{fig:single_pe}
\end{figure}

\newpage\section*{Figure \ref{fig:single_pe}, R.Suda et al., NIM-A} 

\begin{figure}[h]
\epsfysize 10cm
\epsfbox{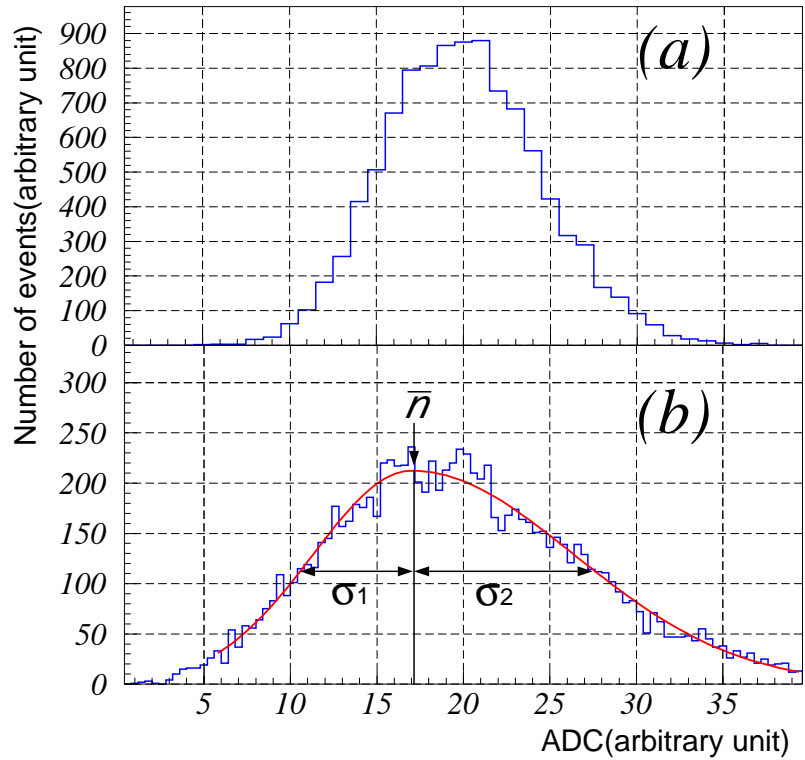}
\caption{Simulated multi-photoelectron spectra: (a) the spectrum of
incident photoelectrons and (b) output photoelectrons.}
\label{fig:deterio_mc}
\end{figure}

\newpage\section*{Figure \ref{fig:deterio_mc}, R.Suda et al., NIM-A} 

\begin{figure}[h]
\epsfysize 10cm
\epsfbox{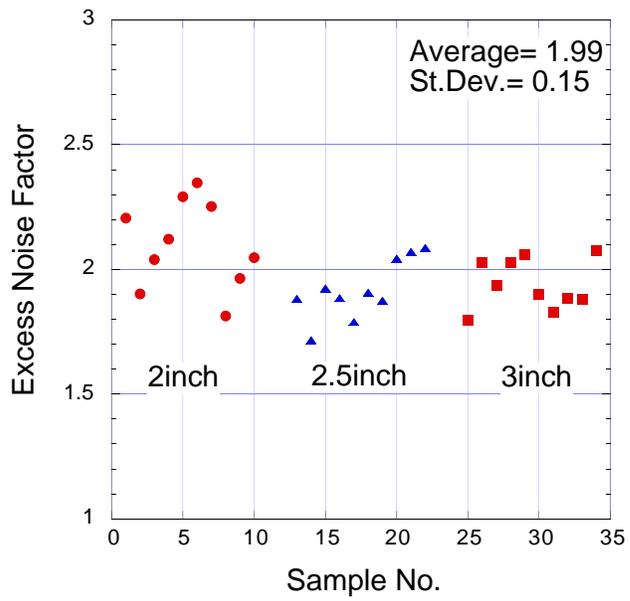}
\caption{Measured Excess Noise Factor
($ENF$) for 2'', 2.5'', and 3'' PMT.}
\label{fig:deterio_exp}
\end{figure}

\newpage\section*{Figure \ref{fig:deterio_exp}, R.Suda et al., NIM-A} 

\begin{figure}[h]
\epsfysize 10cm
\epsfbox{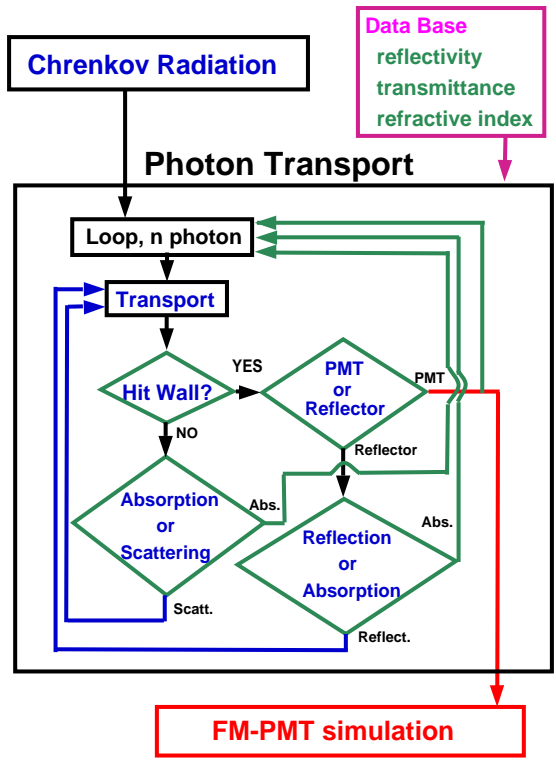}
\caption{Algorithm of the photon transport
part of the Monte-Carlo (ACC++).}
\label{fig:algorithm}
\end{figure}

\newpage\section*{Figure \ref{fig:algorithm}, R.Suda et al., NIM-A} 

\begin{figure}[h]
\epsfbox{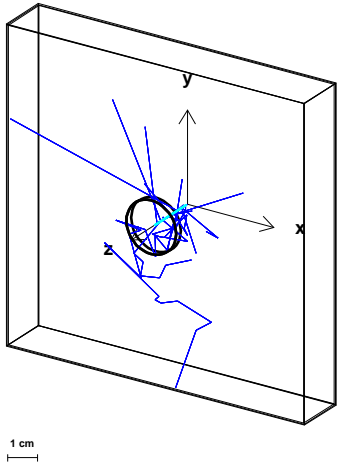}
\caption{Event display of the spectrophotometer simulation.
The lines are the photon trajectories. The circle is the aperture of the 
photodetector.}
\label{fig:spectro}
\end{figure}

\newpage\section*{Figure \ref{fig:spectro}, R.Suda et al., NIM-A} 

\begin{figure}[h]
\epsfysize 10cm
\epsfbox{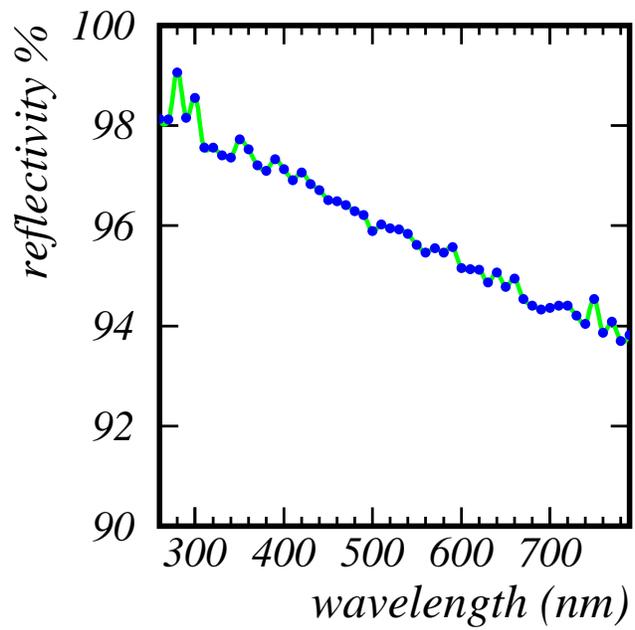}
\caption{Reflectivity of the Goretex as a function of the wavelength
          obtained by the spectrophotometer.}
\label{fig:goretex_ref}
\end{figure}

\newpage\section*{Figure \ref{fig:goretex_ref}, R.Suda et al., NIM-A} 

\begin{figure}[h]
\epsfysize 10cm
\epsfbox{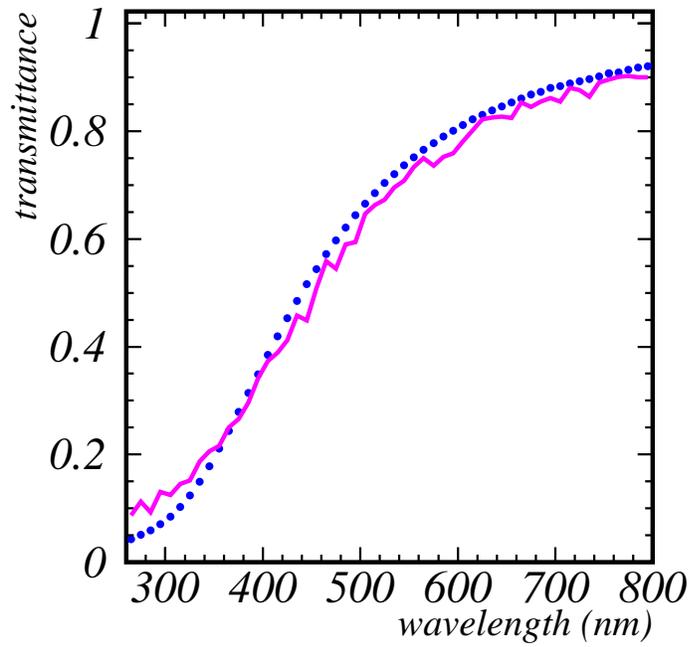}
\caption{Data for the transmittance of n = 1.010 aerogel
(2.4cm thick). The solid
line is the simulation result
with only Rayleigh scattering. }
\label{fig:aperture}
\end{figure}

\newpage\section*{Figure \ref{fig:aperture}, R.Suda et al., NIM-A} 

\begin{figure}[h]
\epsfysize 10cm
\epsfbox{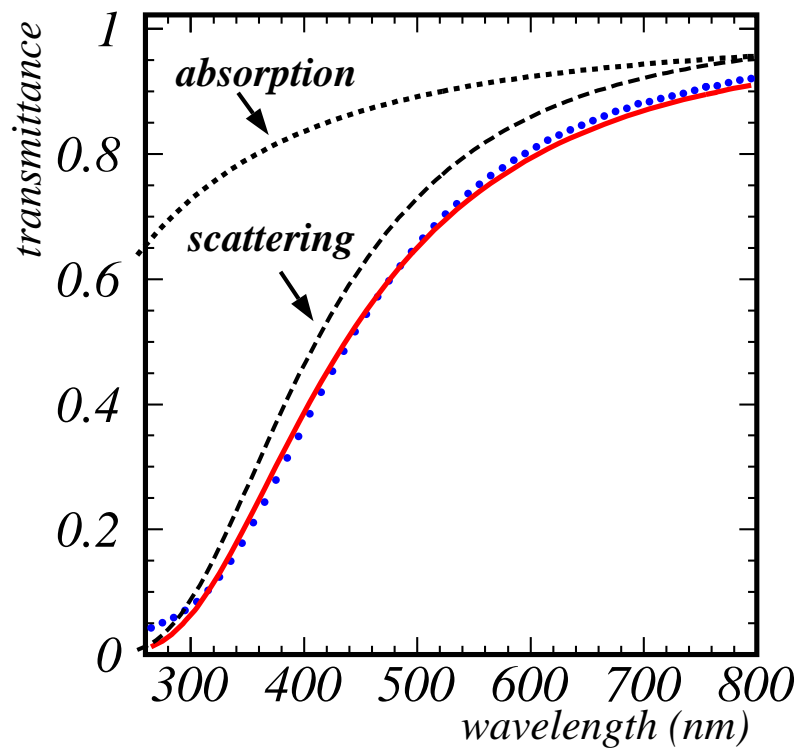}
\caption{Transmittance(circles)
measured by the spectrophotometer for n=1.010, 2cm-thick aerogel;
the solid curve is the best fit with the parameterization
(Equation \ref{eq:slac}),
the dotted line contribution of absorption, and the dashed line
that of scattering.
}
\label{fig:trans_fit}
\end{figure}

\newpage\section*{Figure \ref{fig:trans_fit}, R.Suda et al., NIM-A} 

\begin{figure}[h]
\epsfbox{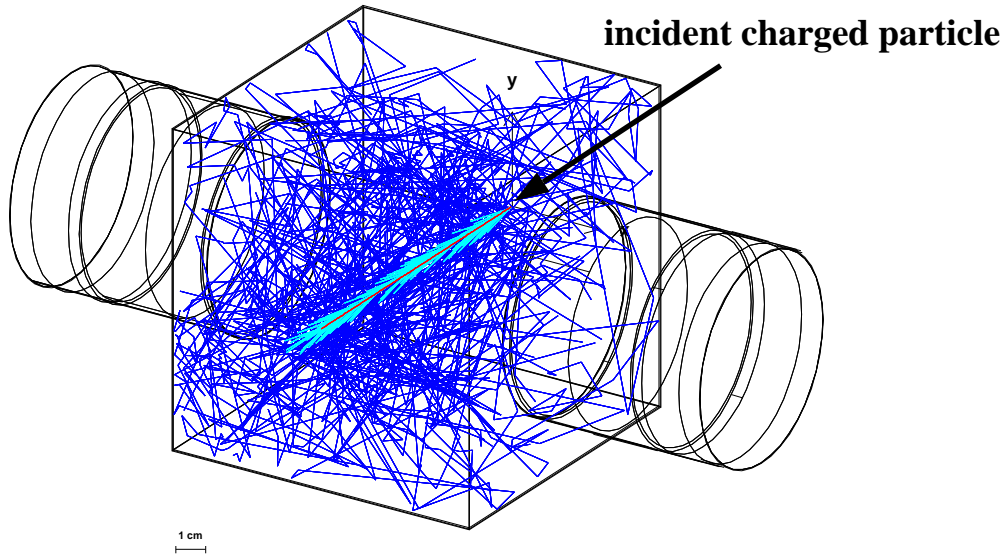}
\caption{Event display of the Monte-Carlo simulation.
The lines are the photon trajectories.}
\label{fig:test}
\end{figure}

\newpage\section*{Figure \ref{fig:test}, R.Suda et al., NIM-A} 

\begin{figure}[h]
\epsfysize 10cm
\epsfbox{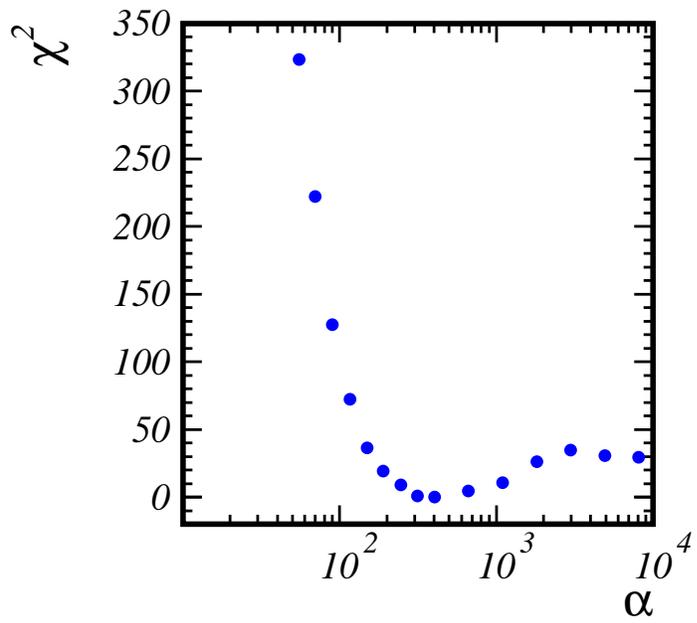}
\caption{$\chi^2$ 
versus $\alpha$. The details of parameter $\alpha$ can be
found in the text.
}
\label{fig:chisq}
\end{figure}

\newpage\section*{Figure \ref{fig:chisq}, R.Suda et al., NIM-A} 

\begin{figure}[h]
\epsfysize 10cm
\epsfbox{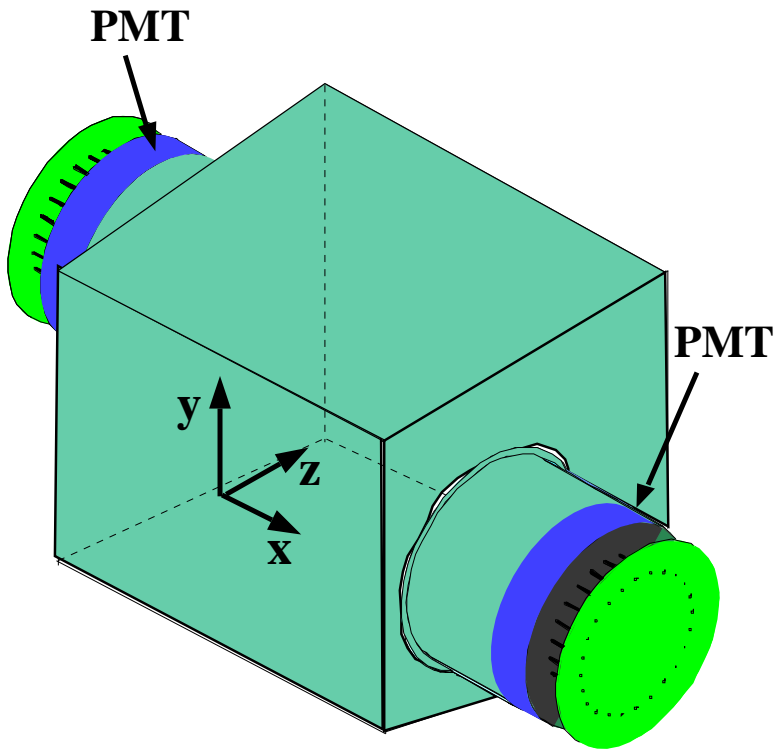}
\caption{Definition of the incident beam position.
The beam direction is parallel to the z-axis.}
\label{fig:pos_inc}
\end{figure}

\newpage\section*{Figure \ref{fig:pos_inc}, R.Suda et al., NIM-A} 

\begin{figure}[h]
\epsfysize 10cm
\epsfbox{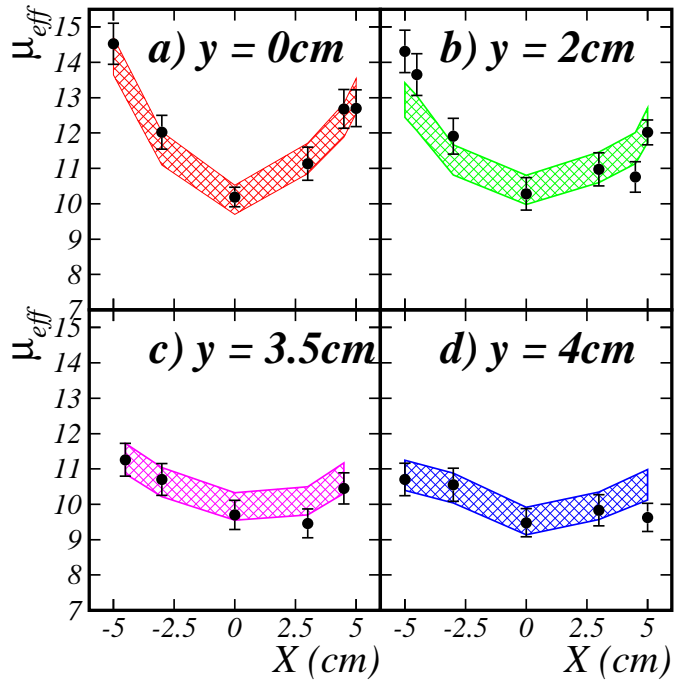}
\caption{Position dependence of $\mu_{eff}$ as a function of 
the incident
beam position: (a) y=0cm, (b) y=2cm, (c) y=3.5cm, and (d) y=4cm.
The points with error bars are the experimental data and the hatched
areas are the Monte-Carlo predictions.
}
\label{fig:pos_dep}
\end{figure}

\newpage\section*{Figure \ref{fig:pos_dep}, R.Suda et al., NIM-A} 

\begin{figure}[h]
\epsfysize 10cm
\epsfbox{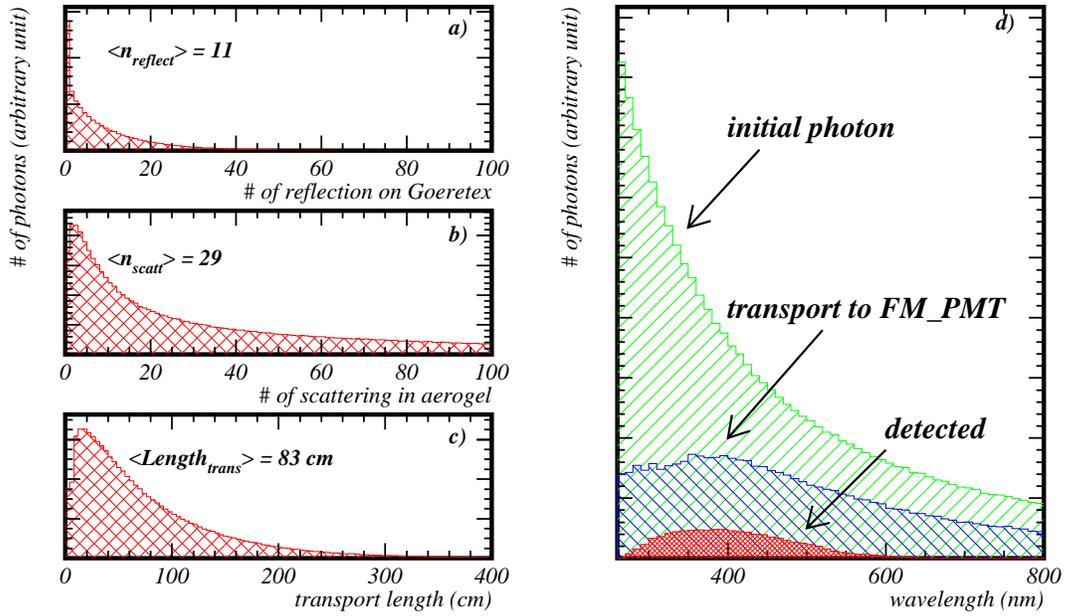}
\caption{Various distributions predicted by the Monte-Carlo simulation:
a) distribution of the number of reflections on Goretex;
b) number of scatterings in aerogel;
c) total transport length inside the counter box; 
and d) wavelength distribution of initial photon (single hatch), 
photon after transport to
FM-PMT(cross hatch), and photon after cut of FM-PMT's quantum
efficiency (dense area).}
\label{fig:trans_tot}
\end{figure}

\newpage\section*{Figure \ref{fig:trans_tot}, R.Suda et al., NIM-A} 

\begin{figure}[h]
\epsfysize 10cm
\epsfbox{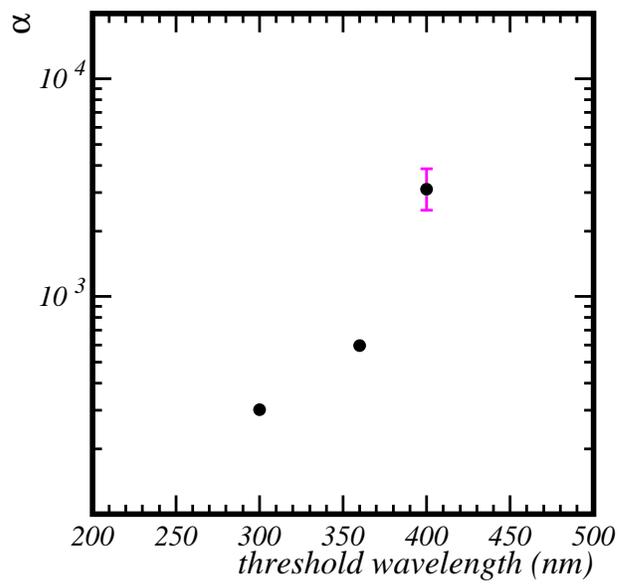}
\caption{Evaluation of the factor $\alpha$ using sharp-cut filters.
The horizontal axis is the threshold wavelength of the filter.
The vertical axis is the
best $\alpha$ values, which were 
obtained by comparing the Monte-Carlo simulation 
data with the beam data.}
\label{fig:filter}
\end{figure}

\newpage\section*{Figure \ref{fig:filter}, R.Suda et al., NIM-A} 

\end{document}